\documentclass{article}
\usepackage{graphicx}
\usepackage{epsfig}

\begin{document}

\begin{center}
{\bfseries Reconstruction of the optical potential in the inverse
quantum scattering. Application to the relativistic inelastic NN
scattering}

\vskip 5mm

N.A. Khokhlov$^{\dag}$, V.A. Knyr$^{\dag \dag}$

\vskip 5mm

{\small  {\it Khabarovsk State University of Technology, 680035
Khabarovsk, Russia }
\\
$\dag$ {\it E-mail: khokhlov@fizika.khstu.ru }; $\dag\dag$ {\it
E-mail: knyr@fizika.khstu.ru }}
\end{center}

\vskip 5mm

\begin{abstract}
The numerical algorithm of the inverse quantum scattering is
developed. This algorithm is based on the Marchenko theory, and
includes three steps. The first one is the algebraic Pad\'{e}
approximation of the unitary S-matrix, what is realized by solving
a system of linear equations. Second step is the exact solution of
the Marchenko equation. The used approximant reduces it to another
system of linear equations. At this step we get the real-valued
potential. It is shown numerically that the developed algorithm is
able not only to generate the given $S$-matrix dependence, but
converges to the initial potential. At third step we construct the
optical complex-valued potential which gives the needed S-matrix.
It is shown that the modern phase shift analysis data allow to
construct the nucleon-nucleon optical potentials of two kinds.
These potentials describe the deuteron properties and the phase
shift analysis data up to 3 GeV and they have different behavior
at short distances. One is a repulsive core potential and another
is a Moscow attractive potential with forbidden states.
\end{abstract}

\vskip 10mm

\section{\textbf{Introduction}}

Quantum inversion has many applications in nuclear physics. Most
of the potential descriptions of the few-body nuclear scattering
base on the inversion of the scattering data. Though
nucleon-nucleon interaction is used as input for all nuclear
calculations, here we always have some fitted parameters and this
fitting is, of course inversion. Moreover all modern
high-precision nucleon-nucleon potentials are fitted to scattering
data and are now perceived as phenomenology. The problem here is
"that quantitative models for the nuclear force have only a poor
theoretical background, while theory based models yield only poor
results" \cite{Mac}. At the same time the nucleon-nucleon phase
shift analysis data are smooth in all investigated energy region
up to 3 GeV \cite{DataScat}. This fact justifies the potential
description without explicit internal degrees of freedom.  It is
commonly supposed that nucleon-nucleon potential is nonlocal.
However local configuration space potentials simplify nuclear
calculations greatly. Nonlocality effects may be treated as
corrections due to the internal degrees of freedom. For these
reasons construction of the high-precision local NN potential
describing at least  NN scattering data is necessary for exact
calculations in nuclear theory. Recently such results appeared in
the literature \cite{Geramb}. To simplify the following
investigations we worked out the presented simple algorithm which
allows to invert scattering data above inelasticity limit and to
get the corresponding optical potential.

  In Sect. 2 we describe our
inversion algorithm, that allows to get the configuration space
potential from phase shift analysis data neglecting inelasticity.
In Sect. 3 we show how to get an optical complex-valued potential
from the real-valued one. This optical potential describes phase
shift analysis data and loss of flux due to inelastic processes.
In Sect. 4 we apply the worked out method to the nucleon-nucleon
scattering data.  We extract two different optical potentials
corresponding to the different asymptotic behavior of the
scattering data.

\section{\textbf{Inversion algorithm}}

 The input data of the Marchenko inversion are
\begin{equation}
\label{eq1} \left\{ {S\left( q \right), \left( { 0 < q < + \infty
} \right),\mbox{  }q_{j},\mbox{  }M_{j},\mbox{  }j =
1,...,n_{\mbox{\tiny{b}}} } \right\}, \label{initial_for_Mar}
\end{equation}

\noindent where $S\left( q \right)$ -- is the scattering matrix
dependance on the momentum $q$, $q^2 = Em$, $q_j ^2 = mE_j \le 0,
\quad E_j $ is the energy of the j-th bound state, so that $\imath
q_j \ge 0$, $m$ -- is the particle (reduced) mass. The $M_j$
matrices give the asymptotic behavior of the corresponding
normalized bound states. These are output data of the partial
scattering Schr\"{o}dinger equation
\begin{equation}
\left[-\frac{d^2}{dr^2}+\frac{l(l+1)}{r^2}+V(r)\right]\psi(r,q)=q^2\psi(r,q)
\label{Schr}
\end{equation}

To illustrate the worked out algorithm first we consider the one
channel case. We proceed from the Marchenko equation
\begin{equation}
\label{eq2} F(x,y) + L(x,y) + \int\limits_x^{ + \infty }
{L(x,t)F(t,y)dt} = 0, \label{Mar}
\end{equation}

\noindent where

\begin{equation}
\label{eq3} F\left( {x,y} \right) = \frac{1}{2\pi }\int\limits_{ -
\infty }^{ + \infty } {h_l^ + \left( {qx} \right)\left( {I -
S\left( q \right)} \right)} h_l^ + \left( {qy} \right)dq +
\sum\limits_{j = 1}^{n_{\mbox{\tiny{b}}} } {M_j^2 } h_l^ + \left(
{iq_j x} \right)h_j^ + \left( {iq_j y} \right),
\end{equation}

\noindent  $h_l^ + \left( z \right)$ are the Riccati-Hankel
functions.

Solution of the eq. (\ref{eq2}) is the function $L\left( {x,y}
\right)$, which gives the reconstructed potential
$V\left(r\right)$ for the eq. (\ref{Schr})

\begin{equation}
\label{eq4} V\left( r \right) = - \frac{dL\left( {r,r}
\right)}{dr}.
\end{equation}

Contrary to the the algorithm used in \cite{Geramb} we use the
algebraic approximant not for the phase shifts $\delta(q)$ but for
the $S$-matrix

\begin{equation}
\label{eq5a} S\left( q \right) =e^{i2\delta}= \frac{f_2 \left( q
\right) - if_1 \left( q \right)}{f_2 \left( q \right) + if_1
\left( q \right)} \quad
\end{equation}
or
\begin{equation}
\label{eq5b}S\left( q \right) = e^{i2\delta}= \left( {\frac{f_2
\left( q \right) - i f_1 \left( q \right)}{f_2 \left( q \right) +
if_1 \left( q \right)}} \right)^2. \end{equation}

In the approximant (\ref{eq5a})  $f_1 \left( q \right)$ and $f_2
\left( q \right)$ are an odd and even polynomials of $q$, which do
not turn to zero at the real axis simultaneously (for $l=0$, in
the case then there is a bound state with $q_1 = 0$, conversely
$f_1 \left( q \right)$ and $f_2 \left( q \right)$ are even and odd
polynomials). In the approximant (\ref{eq5b}) $f_1 \left( q
\right)$ and $f_2 \left( q \right)$ must be of different parity,
but we cannot describe the case with zero energy bound state.  The
important feature of approximants (\ref{eq5a}) and (\ref{eq5b}) is
that these are the most common Pad\'{e} approximants for the
$S$-matrix consistent with its properties. Both approximants lead
to the finite-dimensional kernel
 $F\left( {x,y} \right)$ of the equation (\ref{eq2}).
The function $L\left( {x,y} \right)$ and potential $V\left( r
\right)$ are expressed through the elementary functions
($\sin(r)$, $\cos(r)$ and powers of $r$), so we get Bargmann
potentials. Choice (\ref{eq5a}) or (\ref{eq5b}) is fixed for
needed (phenomenological) dependance $S\left( q \right)$ by
numerical experiment. We choose the relation  which approximates
$S\left( q \right)$ better with less number of  $S\left(q\right)$
poles (taking into account their multiplicity). The less number of
these poles gives the more simple potential.

The properties of the Pad\'{e} approximants are well known, and
there are broad enough functional classes, for which this
approximation converges everywhere besides poles of these
functions. In particular, the consequence of the
Pad\'{e}-hypothesis \cite{Pade} is that if a function is
analytical at point $q=0$ and meromorphic at circle $D$,
containing this point, then its diagonal Pad\'{e} approximations
$[M \mathord{\left/ {\vphantom {M M}} \right.
\kern-\nulldelimiterspace} M]$ (where $M$ is the number of the
approximation poles) converges to this function on compact subsets
of  $D$, which do not contain the poles of this function.
Obviously for the major part of nuclear physics problems the
$S$-matrix does satisfy these conditions. Particularly for the
short range potentials the $S$-matrix is meromorohic at all
complex plane.

Another remark concerns the $S$-matrix poles positions. Our
approximants for $S$-matrix (\ref{eq5a}), (\ref{eq5b}) suggest
that $S$-matrix poles may be positioned everywhere besides the
real axis. Previously at numerical approaches the following
$S$-matrix approximant was used \cite{Wiesner}:

\[
S\left( q \right) = \prod\limits_{\nu = 1}^N {\frac{q - i\alpha
_\nu }{q + i\alpha _\nu }} ,
\]

\noindent with real values $a_\nu $, what assumes that $S$-matrix
poles are positioned only at the imaginary axis. It is obvious,
our approximants are more common and allow to reconstruct the
potentials of more complex form.

The approximants (\ref{eq5a}) and (\ref{eq5b}) lead to the
following expressions for the phase shifts $\delta \left( q
\right)$

\begin{equation}
\label{eq6a} tg\left( { - \delta \left( q \right)} \right) =
\frac{f_1 \left( q \right)}{f_2 \left( q \right)}
\end{equation}
or
\begin{equation}
\label{eq6b} tg\left( { - \frac{1}{2}\delta \left( q \right)}
\right) = \frac{f_1 \left( q \right)}{f_2 \left( q \right)}.
\end{equation}

Having the scattering data set $\delta \left( {q_i } \right) =
\delta _i $ ($i = 1,...,N)$, we get system of $N$ linear equations
which define $N$ unknown coefficients of $f_2 \left( q \right) +
if_1 \left( q \right)$ from (\ref{eq6a}) and (\ref{eq6b}). The
number of experimental phase shifts differs generally from the
chosen number $N$. Therefore the phenomenological dependance
$\delta \left( q \right)$ must be approximated by appropriate
spline. The
 $N$ needed values $\delta \left( {q_i } \right) = \delta _i$ are defined by  some selection from this
 spline.

The increase in number of precise experimental phase shifts would
make approximants (\ref{eq5a}), (\ref{eq5b}) more close to the
true $S$-matrix. The scattering data are defined experimentally
and they have errors. The arbitrary selection of $\delta \left(
{q_i } \right) = \delta _i $ ($i = 1,...,N)$ from the
experimentally defined allowable region may lead to close roots of
the polynomials $f_1 \left( q \right)$ and $f_2 \left( q \right)$,
i.e. to false $S$-matrix poles. Therefore $f_1 \left( q \right)$
and $f_2 \left( q \right)$ must be factored and close roots
particularly those situated near the real axis must be eliminated.
This elimination corresponds to elimination of the Pad\'{e}
approximant defects.

Furthermore, using the approximant (\ref{eq5a}) (in case of
(\ref{eq5b}) the solution is a particular case of the considered
below solution for two bound channels) and calculating the
integral in the eq. (\ref{eq3}) using the residue theorem, we get
the following expression for the kernel $F\left( {x,y} \right)$ of
eq. (\ref{eq1})

\begin{equation}
\label{eq7} F\left( {x,y} \right) = \sum\limits_{i =
1}^{n_{\mbox{\tiny{b}}} } {M_i^2 } h_l^ + \left( {q_i x}
\right)h_l^ + \left( {q_i y} \right) + \sum\limits_{i =
1}^{n_{\mbox{\tiny{pos}}} } {b_i h_l^ + \left( {\beta _i x}
\right)h_l^ + \left( {\beta _i y} \right)} = \sum\limits_{j = 1}^n
{b_j h_l^ + \left( {\beta _j x} \right)h_l^ + \left( {\beta _j y}
\right)} ,
\end{equation}

\noindent where $\beta _i $ ($i = 1,...,n_{\mbox{\tiny{pos}}} )$
are the $S$-matrix poles. Summing is taken over all the $S$-matrix
poles $n_{\mbox{\tiny{pos}}} $ positioned above the real axis and
over bound states  $n_{\mbox{\tiny{b}}} $: $n =
n_{\mbox{\tiny{b}}} + n_{\mbox{\tiny{pos}}} $.

So in this case the kernel of the eq. (\ref{eq2}) is
finite-dimensional. It is known that solution of such equations
reduces to the linear equations solving. We present the specific
solution of this equation for our case.

We search the solution of eq. (\ref{eq2}) in the form

\begin{equation}
\label{eq8} L\left( {x,y} \right) = \sum\limits_{i = 1}^n {P_i
\left( x \right)h_l^ + \left( {\beta _i y} \right)} ,
\end{equation}

\noindent where $P_i \left( x \right)$ are unknown coefficients.
Substituting (\ref{eq7}) and (\ref{eq8}) into (\ref{eq2}) we get
the following  system of linear equations

\begin{equation}
\label{eq9}\sum\limits_{k = 1}^n {A_{ik} } \left( x \right)P_k
\left( x \right) = D_i\left( x \right)\ \ \ \ \ \ (i = 1,..,n),
\end{equation}

\noindent where \begin{equation} \label{eq101}A_{ik} = \delta
_{ik} - b_i \frac{\beta _i h_{l - 1}^ + \left( {\beta _i x}
\right)h_l^ + \left( {\beta _k x} \right) - \beta _k h_l^{+}
\left( {\beta _i x} \right)h_{l - 1}^ + \left( {\beta _k x}
\right)}{\beta _i^2 - \beta _k^2 },\ \ \ \  D_i \left( x \right) =
- b_i h_l^ + \left( {\beta _i x} \right). \end{equation}

The functional coefficients $P_i \left( x \right)$ are defined by
 (\ref{eq9})

\begin{equation}
\label{eq9a} P_i \left( x \right) = \left( {A^{ - 1}D} \right)_i ,
\label{final_Mar}
\end{equation}

\noindent then $L\left( {x,y} \right)$ and the potential $V\left(
r \right)$ from (\ref{eq8}) and (\ref{eq4}).

The worked out technique of the inversion problem was approved on
restoration of the square pit potentials. The accuracy of the
results is illustrated by initial and restored potentials in
fig.1. The convergence of the restoration algorithm with
increasing of power N of the $[N \mathord{\left/ {\vphantom {N
{N]}}} \right. \kern-\nulldelimiterspace} {N]}$ Pad\'{e}
approximant of the $S$-matrix is also shown here.

In case of two bound channels the system of the partial
Schr\"{o}dinger equations is

\begin{equation}
\label{eq10} \left( {\frac{d^2}{dr^2} + V\left( r \right) + \left(
{{\begin{array}{*{20}c}
 {\frac{l_1 \left( {l_1 + 1} \right)}{r^2}} \hfill & 0 \hfill \\
 0 \hfill & {\frac{l_2 \left( {l_2 + 1} \right)}{r^2}} \hfill \\
\end{array} }} \right)} \right)\left( {{\begin{array}{*{20}c}
 {\chi _1 (r) } \hfill \\
 {\chi _2 (r)} \hfill \\
\end{array} }} \right) = q^2\left( {{\begin{array}{*{20}c}
 {\chi _1 (r)} \hfill \\
 {\chi _2 (r)} \hfill \\
\end{array} }} \right),
\end{equation}

\begin{equation}
\label{eq11} V\left( r \right) = \left( {{\begin{array}{*{20}c}
 {V_1 \left( r \right)} \hfill & {V_T \left( r \right)} \hfill \\
 {V_T \left( r \right)} \hfill & {V_2 \left( r \right)} \hfill \\
\end{array} }} \right) \quad ,
\end{equation}

\noindent where $V_1 \left( r \right)$, $V_2 \left( r \right)$ are
potentials in channels 1 and 2, $V_T \left( r \right)$ is
potential bounding them, $\chi _1 (r)$ and $\chi _2 (r)$ are
channel wave functions.

By analogy with  (\ref{eq5b}) we approximate the $S$-matrix by the
following expression

\begin{equation}
\label{eq12}
\begin{array}{l}
 S(x) = \left( {{\begin{array}{*{20}c}
 {\exp \left( {2i\delta _1 } \right)\cos 2\varepsilon } \hfill & {i\exp
\left( {i\left( {\delta _1 + \delta _2 } \right)} \right)\sin
2\varepsilon }
\hfill \\
 {i\exp \left( {i\left( {\delta _1 + \delta _2 } \right)} \right)\sin
2\varepsilon } \hfill & {\exp \left( {2i\delta _2 } \right)\cos
2\varepsilon
} \hfill \\
\end{array} }} \right) = \\
 = \left( {{\begin{array}{*{20}c}
 {\left( {\frac{f_2^{\left( 1 \right)} \left( q \right) - if_1^{\left( 1
\right)} \left( q \right)}{f_2^{\left( 1 \right)} \left( q \right)
+ i f_1^{\left( 1 \right)} \left( q \right)}}
\right)^2\frac{\left( {f_2^{\left( {12} \right)} \left( q \right)}
\right)^2 - \left( {f_1^{\left( {12} \right)} \left( q \right)}
\right)^2}{\left( {f_2^{\left( {12} \right)} \left( q \right)}
\right)^2 + \left( {f_1^{\left( {12} \right)} \left( q \right)}
\right)^2}} \hfill & { - 2 i \frac{f_2^{\left( {12} \right)}
\left( x \right)f_1^{\left( {12} \right)} \left( x \right)}{\left(
{f_2^{\left( {12} \right)} \left( q \right)} \right)^2 + \left(
{f_1^{\left( {12} \right)} \left( q \right)}
\right)^2}\prod\limits_{j = 1,2} {\frac{f_2^{\left( j \right)}
\left( q \right) - if_1^{\left( j \right)} \left( q
\right)}{f_2^{\left( j \right)} \left( q \right) + if_1^{\left( j
\right)} \left( q \right)}} } \hfill \\
 { - 2i \frac{f_2^{\left( {12} \right)} \left( q \right)f_1^{\left( {12}
\right)} \left( q \right)}{\left( {f_2^{\left( {12} \right)}
\left( q \right)} \right)^2 + \left( {f_1^{\left( {12} \right)}
\left( q \right)} \right)^2}\prod\limits_{j = 1,2}
{\frac{f_2^{\left( j \right)} \left( q \right) - i f_1^{\left( j
\right)} \left( q \right)}{f_2^{\left( j \right)} \left( q \right)
+ i f_1^{\left( j \right)} \left( q \right)}} } \hfill & {\left(
{\frac{f_2^{\left( 2 \right)} \left( q \right) - i f_1^{\left( 2
\right)} \left( q \right)}{f_2^{\left( 2 \right)} \left( q \right)
+ i f_1^{\left( 2 \right)} \left( q \right)}}
\right)^2\frac{\left( {f_2^{\left( {12} \right)} \left( q \right)}
\right)^2 - \left( {f_1^{\left( {12} \right)} \left( q \right)}
\right)^2}{\left( {f_2^{\left( {12} \right)} \left( q \right)}
\right)^2 + \left( {f_1^{\left( {12} \right)} \left( q
\right)} \right)^2}} \hfill \\
\end{array} }} \right) \\
 \end{array}
\end{equation}

Here we again choose the most general form of Pad\'{e} approximant
in contrast to the form used in \cite{Wiesner}. The positions of
$S$-matrix poles are determined from the equations analogous to
(\ref{eq6b}), but with additional equations for the mixing
parameter

\begin{equation}
\label{eq13} tg\left( {\varepsilon \left( q \right)} \right) =
\frac{f_1 \left( q \right)}{f_2 \left( q \right)},
\end{equation}
with scattering data $\varepsilon \left( q_i
\right)=\varepsilon_i$.

In case of bound channels and $l_i\neq 0$, in the original
Marchenko theory   it was proposed to use transformation of the
initial eqs.  (\ref{eq10}) with $l_1 = l_2 = 0$. The same approach
was used in \cite{Wiesner}. Nevertheless, from the numerical point
of view this approach is less effective than the direct solution
of the inverse problem with generalized Marchenko equation
\cite{Blaz}. Formally it has the former view

\begin{equation}
\label{eq14} L\left( {x,y} \right) + F\left( {x,y} \right) +
\int\limits_x^{ + \infty } {L\left( {x,t} \right)F\left( {t,y}
\right)dt} = 0,
\end{equation}

\noindent but functions involved are matrices $\left( {2\times 2}
\right)$

\begin{equation}
\label{eq15} F\left( {x,y} \right) = \frac{1}{2\pi }\int\limits_{
- \infty }^{ + \infty } {H\left( {qx} \right)\left[ {1 - S\left( q
\right)} \right]H\left( {qy} \right)dq} + \sum\limits_{i =
1}^{n_{b} } {H\left( {\beta _i x} \right)M_i H\left( {\beta _i y}
\right)} ,
\end{equation}

\noindent where

\begin{equation}
\label{eq16} H\left( x \right) = \left( {{\begin{array}{*{20}c}
 {h_{l_1 }^ + \left( x \right)} \hfill & 0 \hfill \\
 0 \hfill & {h_{l_2 }^ + \left( x \right)} \hfill \\
\end{array} }} \right) \, .
\end{equation}

Using the chosen $S$-matrix approximant and the residue theorem
 we get the following expression for $F\left( {x,y} \right)$

\begin{equation}
\label{eq17}
\begin{array}{l}
 F\left( {x,y} \right) = i\sum\limits_{Im\beta _i > 0} Res H\left(qx\right)\left( {I - S\left( q \right)} \right)H\left( {qy} \right) +
\sum\limits_{i = 1}^{n_{\mbox{\tiny{b}}} } {H\left( {\beta _i x}
\right)M_i H\left(
{\beta _i y} \right)} = \\
 = \sum\limits_{Im\beta _i > 0} {H\left( {\beta _i x} \right)Q_i^1 H\left(
{\beta _i y} \right)} + \sum\limits_{\beta _i \in A} {x{H}'\left(
{\beta _i
x} \right)Q_i^2 H\left( {\beta _i y} \right)} + \\
 + \sum\limits_{\beta _i \in \textbf{A}} {H\left( {\beta _i x} \right)Q_i^2
{H}'\left( {\beta _i y} \right)y} , \\
 \end{array}
\end{equation}

\noindent here $Q_i^j\ \ \left( {j = 1,2} \right)$ are constant
matrices,

\[
{H}'\left( x \right) = \left( {{\begin{array}{*{20}c}
 {{dh_{l_1 }^ + \left( x \right)} \mathord{\left/ {\vphantom {{dh_{l_1 }^ +
\left( x \right)} {dx}}} \right. \kern-\nulldelimiterspace} {dx}}
\hfill & 0
\hfill \\
 0 \hfill & {{dh_{l_2 }^ + \left( x \right)} \mathord{\left/ {\vphantom
{{dh_{l_2 }^ + \left( x \right)} {dx}}} \right.
\kern-\nulldelimiterspace}
{dx}} \hfill \\
\end{array} }} \right),
\]

\noindent $\beta _i $ are the $S$-matrix poles, values $q_{j}$
correspond to bound states, $\textbf{A}$ is the set of $S$-matrix
poles of the second order positioned above the real axis.

We solve eq. (\ref{eq14}) using substitution

\begin{equation}
\label{eq18} L\left( {x,y} \right) = \sum\limits_{i = 1}^n {P_i
\left( x \right)} H\left( {\beta _i y} \right) + \sum\limits_{i =
1}^n {N_i \left( x \right)} y{H}'\left( {\beta _i y} \right),
\end{equation}

\noindent where summing is taken over all the $S$-matrix poles
$n_{\mbox{\tiny{pos}}} $ positioned above the real axis and over
all values $q_{j}$. This construction leads to the system of
linear equations for the functional  $\left( {2\times 2} \right)$
matrix-coefficients  $P_i \left( x \right)$, $N_i \left( x
\right)$

\begin{equation}
\label{eq18b}\begin{array}{r}
 \sum\limits_i {P_i \left( x \right)Q_{ij}^3 \left( x \right)} +
\sum\limits_i {N_i \left( x \right)} Q_{ij}^5 \left( x \right) =
H\left(
{\beta _j x} \right)Q_j^1 + x{H}'\left( {\beta _j x} \right)Q_j^2 \\
 \sum\limits_i {N_i \left( x \right)Q_{ij}^6 \left( x \right)} +
\sum\limits_i {P_i \left( x \right)} Q_{ij}^4 \left( x \right) =
H\left(
{\beta _j x} \right)Q_j^2 \\
 \end{array}\end{equation}

\noindent where

\[
Q_{ij}^3 \left( x \right) = I\delta _{ij} + \int\limits_x^{ +
\infty } {H\left( {\beta _i t} \right)H\left( {\beta _j t}
\right)dt\times Q_j^1 } + \int\limits_x^{ + \infty } {tH\left(
{\beta _i t} \right){H}'\left( {\beta _j t} \right)dt\times Q_j^2
}
\]

\begin{equation}
\label{eq19} Q_{ij}^4 \left( x \right) = \int\limits_x^{ + \infty
} {H\left( {\beta _i t} \right)H\left( {\beta _j t}
\right)dt\times Q_j^2 }
\end{equation}

\[
Q_{ij}^5 \left( x \right) = \int\limits_x^{ + \infty }
{t{H}'\left( {\beta _i t} \right)H\left( {\beta _j t}
\right)dt\times Q_j^1 } + \int\limits_x^{ + \infty }
{t^2{H}'\left( {\beta _i t} \right){H}'\left( {\beta _j t}
\right)dt\times Q_j^2 }
\]

\[
Q_{ij}^6 \left( x \right) = I\delta _{ij} + \int\limits_x^{ +
\infty } {t{H}'\left( {\beta _i t} \right)H\left( {\beta _j t}
\right)dt\times Q_j^2 } ,
\]

\noindent $I$ is the unit matrix. The integrals in expressions
(\ref{eq19}) are easily calculated analytically, but are
cumbersome so we do not present them.

Having solved this linear equation system we get the sought-for
potential from (\ref{eq18}) and (\ref{eq4}).

The multichannel generalization is made analogously.

\section{The optical potential}
\mbox{}

The Marchenko inversion does not give the needed optical
potential. But found potential may serve as the initial potential
for the iteration procedure that converts it into the optical
(complex-valued) potential that describes inelastic processes.

First we consider the one channel problem.

The phase equation \cite{Calo} for the initial potential
$V^0\left( r \right)$ obtained by some inversion procedure (say
from Marchenko equation) is

\begin{equation}
\label{eq20} \delta _l^{(0)} = - \frac{1}{k}\int\limits_0^\infty
{V^{(0)}\left( r \right)D_l^2 \left( {qr} \right)\sin ^2\left( {qr
+ \delta ^{(0)}\left( r \right)} \right)dr} ,
\end{equation}

\noindent where  $D_l \left( z \right)$ is Riccati-Bessel
amplitude \cite{Calo}.

Let us consider the complex-valued potential $V^{(1)}\left( r
\right)$ obtained from $V^0\left( r \right)$ by transformation

\begin{equation}
\label{eq21} V^{(1)}\left( r \right) = \left( {1 + i\alpha }
\right)V^{(0)}\left( r \right),
\end{equation}

\noindent where $\alpha $ is some real parameter. Evidently the
phase equation for this potential is

\begin{equation}
\label{eq22} \delta ^{(1)} = - \frac{1}{k}\left( {1 + i\alpha }
\right)\int\limits_0^\infty {V^{(0)}\left( r \right)D_l^2 \left(
{qr} \right)\sin ^2\left( {qr + \delta ^{(1)}\left( r \right)}
\right)dr} .
\end{equation}

From eqs. (\ref{eq20}) and (\ref{eq22}) we get

\begin{equation}
\label{eq23}
\begin{array}{l}
 \delta ^{(1)} - \left( {1 + i\alpha } \right)\delta ^{(0)} = \\
 = - \frac{1 + i\alpha }{k} \int\limits_0^\infty
{V^{(0)}\left( r \right)D_l^2 \left( {kr} \right)\left( {\sin
^2\left( {qr + \delta ^{(1)}\left( r \right)} \right) - \sin
^2\left( {qr + \delta
^{(0)}\left( r \right)} \right)} \right)dr} = \\
 = - \frac{1 + i\alpha }{k}\int\limits_0^\infty
{V^{(0)}\left( r \right)D_l^2 \left( {kr} \right)\sin \left( {2qr
+ \delta ^{(1)}\left( r \right) + \delta ^{(0)}\left( r \right)}
\right)\sin \left( {\delta ^{(1)}\left( r \right) - \delta
^{(0)}\left( r \right)} \right)dr}
\\
 \end{array}
\end{equation}

For smooth enough potentials and  $\alpha \equiv \alpha \left( q
\right)$ not rapidly increasing with increasing of  $q$ the right
side of eq. (\ref{eq23}) rapidly decreases comparing with
 $\delta ^{(0)}$ and $\delta ^{(1)}$, because under the integral in (\ref{eq23})
 there is more rapidly oscillating function than in   (\ref{eq20}) and (\ref{eq22}).
 Then as a first approximation we may take

\begin{equation}
\label{eq24} \delta ^{(1)} \approx \left( {1 + i\alpha }
\right)\delta ^{(0)} = \delta _R + i\delta _I .
\end{equation}

For inelastic scattering the $S$-matrix is expressed through the
real inelastic parameter  $\rho $ and the real phase shift $\delta
$

\begin{equation}
\label{eq25} S = \cos ^2\left( \rho \right)e^{2i\delta } =
e^{2i\left( {\delta _R + i\delta _I } \right)},
\end{equation}

\noindent so it is easily arrived at

\begin{equation}
\label{eq26} \delta \approx \delta ^{(0)},
\end{equation}

\begin{equation}
\label{eq27} \cos ^2\left( \rho \right) \approx e^{ - 2\alpha
\delta ^{(0)}},
\end{equation}

\noindent whence it follows that $\alpha \delta \ge 0$.

The formula (\ref{eq27}) allows to calculate the parameter $\alpha
$ from the known values $\rho $ and $\delta ^{(0)} \approx \delta
$.

This approximation works well enough for many quantum mechanical
problems. For example in case of nucleon-nucleon scattering $\rho
= 0$ below the inelasticity limit which is high enough and $\rho $
grows slowly enough above this limit.

Nevertheless in some instances the initial potential
$V^{(0)}\left( r \right)$ must be corrected. In this case the
following iteration procedure may be efficient. First from the
formula (\ref{eq27}) values $\alpha(q)$ are calculated from
$\rho(q)$ and $\delta(q)$. Then with the potential $V^{(1)}\left(
r \right)$, defined by formula (\ref{eq21}) new phase shifts
$\delta ^{\left( 1 \right)}$ are calculated from (\ref{eq22}). If
they do not satisfy phase shift analysis data  $\delta ^{(0)}\pm
\Delta \delta $ then changing the input data to $\delta
^{(\ref{eq2})} = 2\delta ^{(0)} - \delta ^{(1)}$ we repeat
inversion (eqs. (\ref{initial_for_Mar})-(\ref{final_Mar})) and
determine new $\alpha $ from formula (\ref{eq27}).

The case of two channels  is considered in a like manner, though
the final expressions are more complex. By analogy with the one
channel case the following generalization for the optical
potential is derived (the mixing parameter $\varepsilon $ is
considered small):

\begin{equation}
\label{eq28} V^{(1)}\left( r \right) = \left(
{{\begin{array}{*{20}c}
 {\left( {1 - i\alpha } \right)V_{11}^{(0)} } \hfill & {\left( {1 - {i\left(
{\alpha + \beta } \right)} \mathord{\left/ {\vphantom {{i\left(
{\alpha + \beta } \right)} 2}} \right. \kern-\nulldelimiterspace}
2}
\right)V_{12}^{(0)} } \hfill \\
 {\left( {1 - {i\left( {\alpha + \beta } \right)} \mathord{\left/ {\vphantom
{{i\left( {\alpha + \beta } \right)} 2}} \right.
\kern-\nulldelimiterspace} 2} \right)V_{12}^{(0)} } \hfill &
{\left( {1 - i\beta } \right)V_{22}^{(0)}
} \hfill \\
\end{array} }} \right)
\end{equation}

The first $S$-matrix approximation is

\begin{equation}
\label{eq29}
\begin{array}{l}
 S^{(1)} = \left( {{\begin{array}{*{20}c}
 {e^{\left( {2i\delta _1 } \right)}\cos 2\varepsilon\cos ^2\rho
_1^{} } \hfill & {e^{i\left( {\delta _1 + \delta _2 } \right)}\sin
2\varepsilon \cos \rho _1^{(1)} \cos \rho _2 }
\hfill \\
 {e^{i\left( {\delta _1 + \delta _2} \right)}\sin 2\varepsilon
^{(1)}\cos \rho _1 \cos \rho _2 } \hfill & {e^{\left( {2i\delta
_2} \right)}\cos 2\varepsilon \cos ^2\rho _2 } \hfill \\
\end{array} }} \right) = \\
 = \left( {{\begin{array}{*{20}c}
 {e^{2i\left( {1 + i\alpha } \right)\delta _1^{(0)} }\cos 2\left( {1 +
i\frac{\left( {\alpha + \beta } \right)}{2}} \right)\varepsilon
^{(0)}} \hfill & {e^{i\left( {\delta _1^{(1)} \left( {1 + i\alpha
} \right) + \delta _2^{(1)} \left( {1 + i\beta } \right)}
\right)}\sin 2\left( {1 + i\frac{\left( {\alpha + \beta }
\right)}{2}} \right)\varepsilon ^{(0)}}
\hfill \\
 {e^{i\left( {\delta _1^{(1)} \left( {1 + i\alpha } \right) + \delta
_2^{(1)} \left( {1 + i\beta } \right)} \right)}\sin 2\left( {1 +
i\frac{\left( {\alpha + \beta } \right)}{2}} \right)\varepsilon
^{(0)}} \hfill & {e^{2i\left( {1 + i\beta } \right)\delta _2^{(1)}
}\cos 2\left( {1 + i\frac{\left( {\alpha + \beta } \right)}{2}}
\right)\varepsilon ^{(0)}}
\hfill \\
\end{array} }} \right) \\
 \end{array}
\quad
\end{equation}

As all the parameters and phase shifts are real, then eq.
(\ref{eq29}) links between parameters $\alpha $ and $\beta $ with
the mixing parameters $\rho _i $  and phase shifts
$\delta_{i}=\delta^{(0)}_{i}$.

This consideration may be generalized to the multichannel case
though calculations become much more complex.

\section{The optical nucleon-nucleon potential }
\mbox{}

We applied  the described algorithm of inversion to reconstruction
 of the nucleon-nucleon potential. As input data for this reconstruction
 we used modern phase shift analysis data  up to  1100 MeV for ${
}^3S_1 - { }^3D_1 $ state and up to 3 GeV for ${ }^1S_0 $ state of
nucleon-nucleon system \cite{DataScat}. The deuteron properties
were taken from \cite{DataDeut}.

 The relativistic effects were
taken into account in frames of relativistic quantum mechanics of
systems with a fixed number of particles (point form dynamics).
The review of this approach can be found in \cite{RQM}.

This approach is based on the assumption that at not high energies
we may consider the number of particles fixed, but the invariance
group is the Poincare group. A system of two particles is
described by the wave function, which is an eigenfunction of the
mass operator or the mass squared operator. In this case we may
represent this wave function as a product of the external and
internal wave functions \cite{Lev,Khokhlov}. The internal wave
function is also an eigenfunction of the mass operator or the mass
squared operator. It is shown that the mass squared method is
consistent with conventional fitting of the Lorentz invariant
cross section as a function of laboratory energy \cite{mass}. We
consider system of two particles (nucleons) with equal masses.
Then the internal wave function
 $\chi _ ({\rm {\bf q}})$ satisfies the following equation

\begin{equation}
\label{eq30} \left[ {4(m^2 + {\rm {\bf q}}^2) + V} \right]\chi =
M^2\chi
\end{equation}
or
\begin{equation}
\label{eq30b}  \left( {\frac{{\rm {\bf q}}^2}{m} + \frac{V}{4m}}
\right)\chi = E\chi ,
\end{equation}
where
\[
  E = \frac{M^2 -
4m^2}{4m} = \frac{\kappa ^2}{m}.
\]

The eq. (\ref{eq30b}) formally coincides with the Schr\^{o}dinger
equation. In eqs. (\ref{eq30}), (\ref{eq30b}) $M^2$ is the mass
squared operator, $m$ is the mass of nucleon, $V$ is the
nucleon-nucleon potential, ${\rm {\bf q}}$ is the momentum
operator of one of the nucleons in the center of masses system. We
use system $\hbar = c = 1$. The quasicoordinate representation
corresponds to the realization  ${\rm {\bf q}} = - i\frac{\partial
}{\partial {\rm {\bf r}}}$, $V=V({\rm {\bf r}})$.  In
\cite{Khokhlov2} we showed that this formalism can be easily
generalized to the case of inelastic channels, particularly it
allows to take into account the isobar channels in NN scattering.

 This formal coincidence allows us to apply our inversion algorithm.
The modern phase shift analysis data \cite{DataScat} allow to
construct nucleon-nucleon potentials of two different kinds,
depending on the asymptotic behavior  of the  phase shifts above
investigated energy region of 3 GeV. We constructed the
nucleon-nucleon optical potentials of two kinds for $^1 S_0$ and
$^3 S_1-^3 D_1$ partial waves.  These potentials describe the
deuteron properties and the phase shift analysis data up to 3 GeV
and they have different behavior at short distances. One is a
repulsive core potential and another is a Moscow attractive
potential with forbidden states. The Moscow potential was
introduced in \cite{Neud}. These potentials are not phase
equivalent and their $S$-matrices differ even bellow 3 GeV, but
within the experimental errors. The $^1 S_0$  phase shift of
Moscow potential begins from $\pi$. $^3 S_0$ phase shifts of
Moscow potential begin from $2\pi$. The mixing parameter
$\epsilon_1$ of Moscow potential differs from this of repulsive
core potential by sign. Above 3 GeV these potentials give
different $S$-matrices.

The real parts of the constructed partial potentials are presented
in fig. 2  and fig. 3. They may be downloaded from
www.physics.khstu.ru in numerical form. Standard notation for
central and tensor parts is used, so for $^3 S_1-^3 D_1$

\[V_{CS}(r)=V_1(r), \,\,\,\,\,
V_{CD}(r)=V_2(r)+\left.V_T(r)\right/2\sqrt{2},\,\,\,\,
V_{tens}(r)=V_T(r)/\sqrt{2}\]

The imaginary parts of potentials are defined by eqs. (\ref{eq21},
\ref{eq28}),
  where parameters $\alpha$ and $\beta$
can be easily calculated from the phase shift analysis data
\cite{DataScat}. The phase shifts and mixing parameter are
compared with the phase shift analysis data \cite{DataScat} in
fig. 4. In table 1 the results of our deuteron properties
calculations are compared with the experimental data
\cite{DataDeut}. Both kinds of the constructed partial potentials
describe the data well in the limits of the experimental errors.

We conclude that the available data of the NN scattering and the
deuteron properties are not discriminative with respect to the
kind of the NN potential. To discriminate between the considered
two kinds of nucleon-nucleon potential we need careful
experimental examination of other inelastic processes such as
$pp\rightarrow pp\gamma$ \cite{Khokhlov} or $^2 H+\gamma
\rightarrow n+p$.

\newpage

\renewcommand{\baselinestretch}{1}
\begin{table}[t]
{Table 1. The deuteron properties}\\
\begin{tabular}
{|p{87pt}|p{87pt}|p{73pt}|p{73pt}|} \hline & Exp. $^a$ & Calculation  & Calculation\\
&  & with repulsive  & with Moscow \\
 &  & core potential   & potential \\          \hline Energy (MeV) & 2,22458900(22)& 2,2246$^a$& 2,2246$^a$ \\
\hline $Q$ (Fm$^{2}$)& 0,2859(3)& 0,2639$^c$ & 0,277$^c$\\
\hline A$_{S}$  (Fm$^{ - 1 / 2})$& 0,8802(20)& 0,8802& 0,8802\\
\hline r$_{d}$ (Fm)& 1,9627(38)& 1,951 & 1,956\\
\hline  $\eta_{d / s} $ & 0,02714 & 0,02714 & 0,02714\\
\hline  $\mu _{d}$& 0,857406(1)& 0,8497$^c$ & 0,859$^c$\\
\hline
\end{tabular}
\label{tab1}\vspace*{0.5cm}\\
$^a$ relativistic correction included; $^b$ Data are from
\cite{DataDeut}; $^c$ Meson exchange currents are not included.
\end{table}

\begin{figure}[b] \epsfysize=50mm \centerline{
\epsfbox{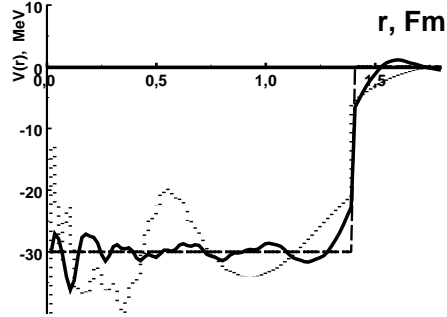}} \caption{Reconstructed square pit potential.
Solid line N=31, dotted line N=23. N is power of the $[N/N]$
Pad\'{e} approximation.}
\end{figure}

\begin{figure}[t] \epsfysize=50mm \centerline{
\epsfbox{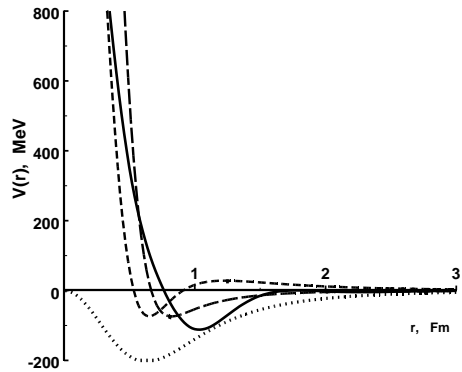}} \caption{Real parts of optic repulsive core
potentials. Solid line $^1 S_0$ -one channel. Two bound channels
long dashed line $V_{CS}(r)$ ($^3 S_1$), short dashed line
$V_{CD}(r)$ ($^3 D_1$), dotted line $V_{tens}(r)$.}
\end{figure}

\begin{figure}[b] \epsfysize=50mm \centerline{
\epsfbox{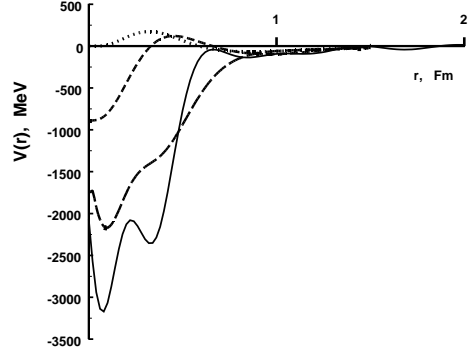}} \caption{Same as in in Fig. 2 but for the Moscow
potentials.}
\end{figure}

\begin{figure}[t]
\epsfysize=50mm \centerline{ \epsfbox{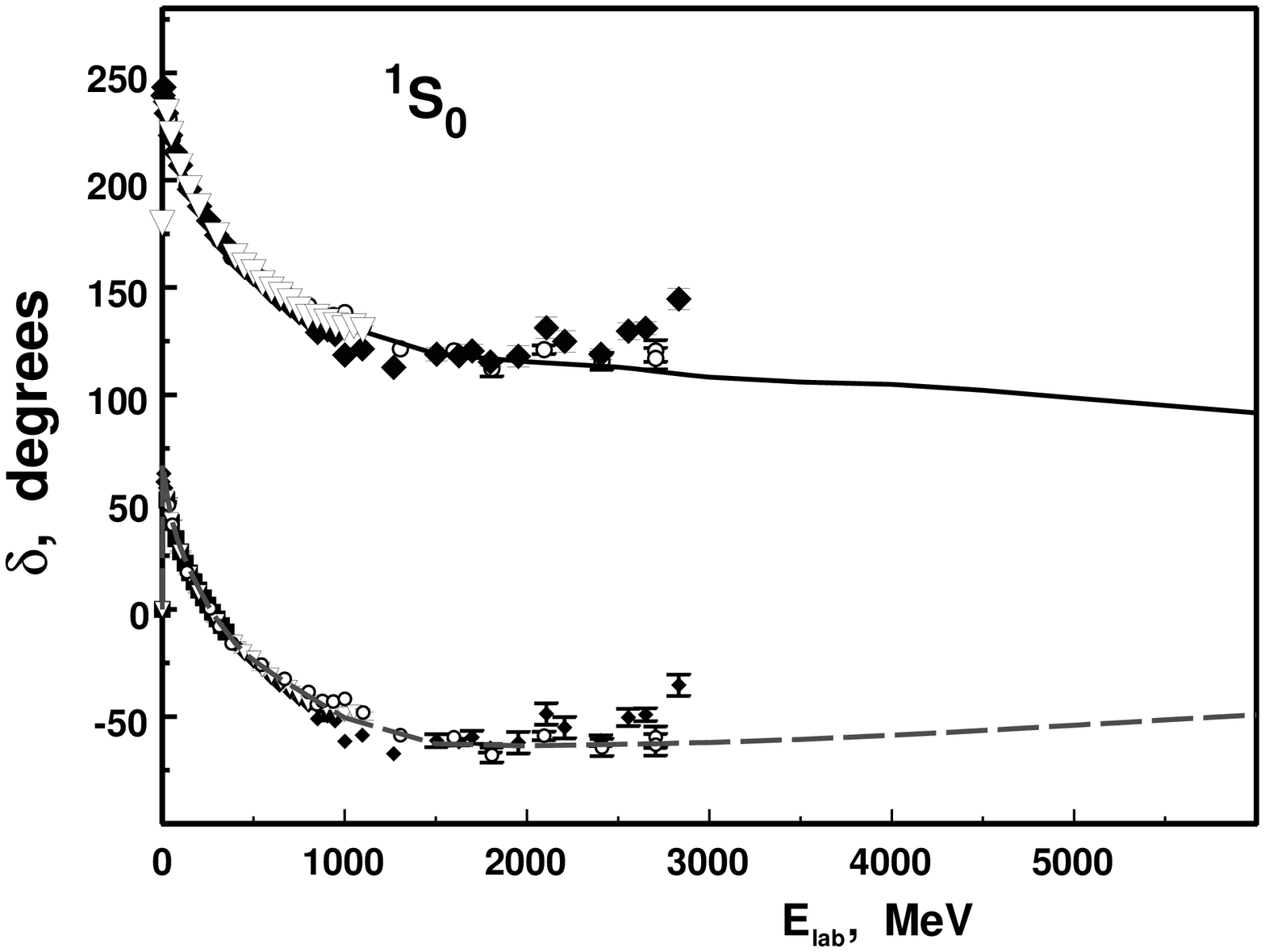} \epsfysize=50mm
\epsfbox{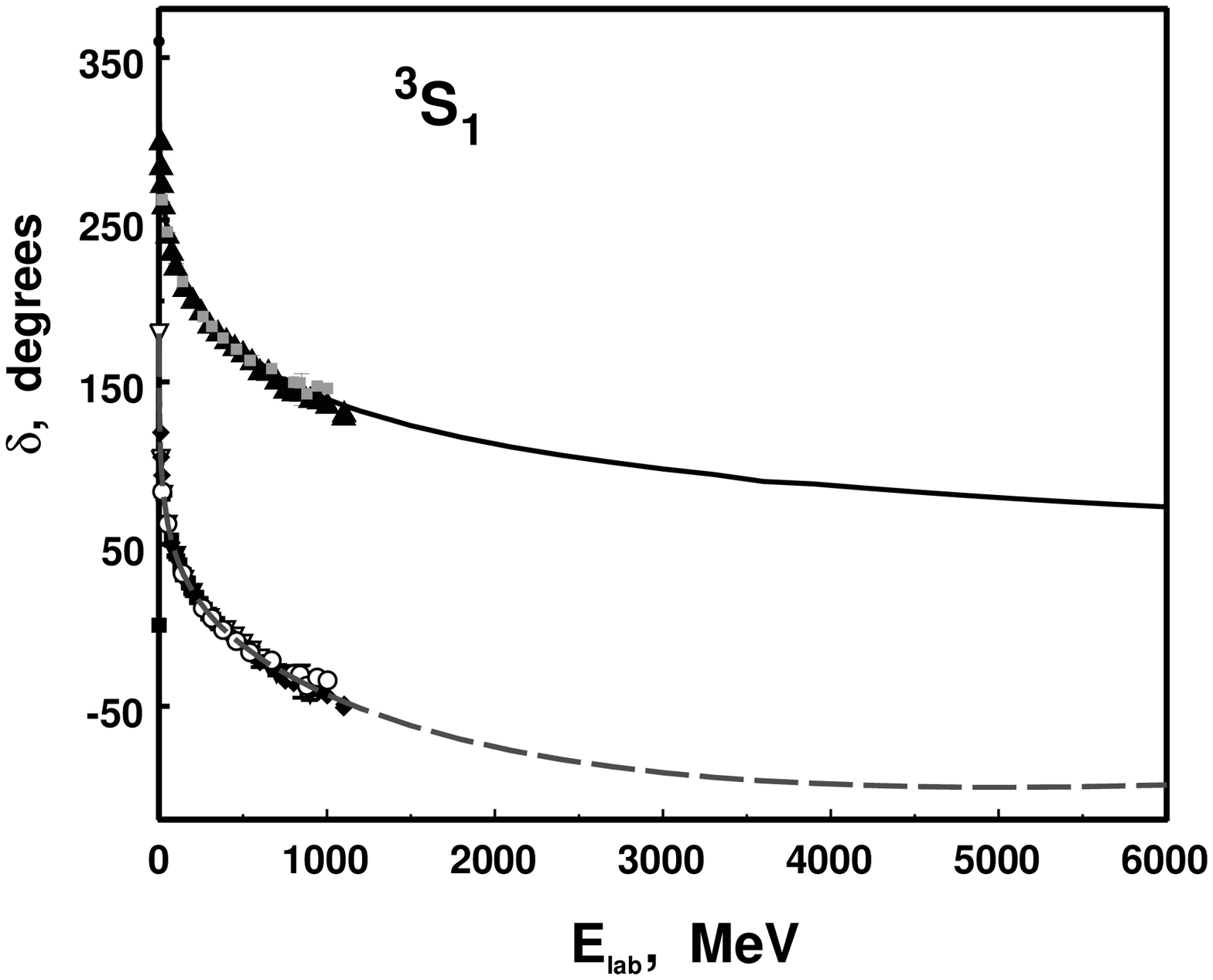}} \epsfysize=50mm \centerline{ \epsfbox{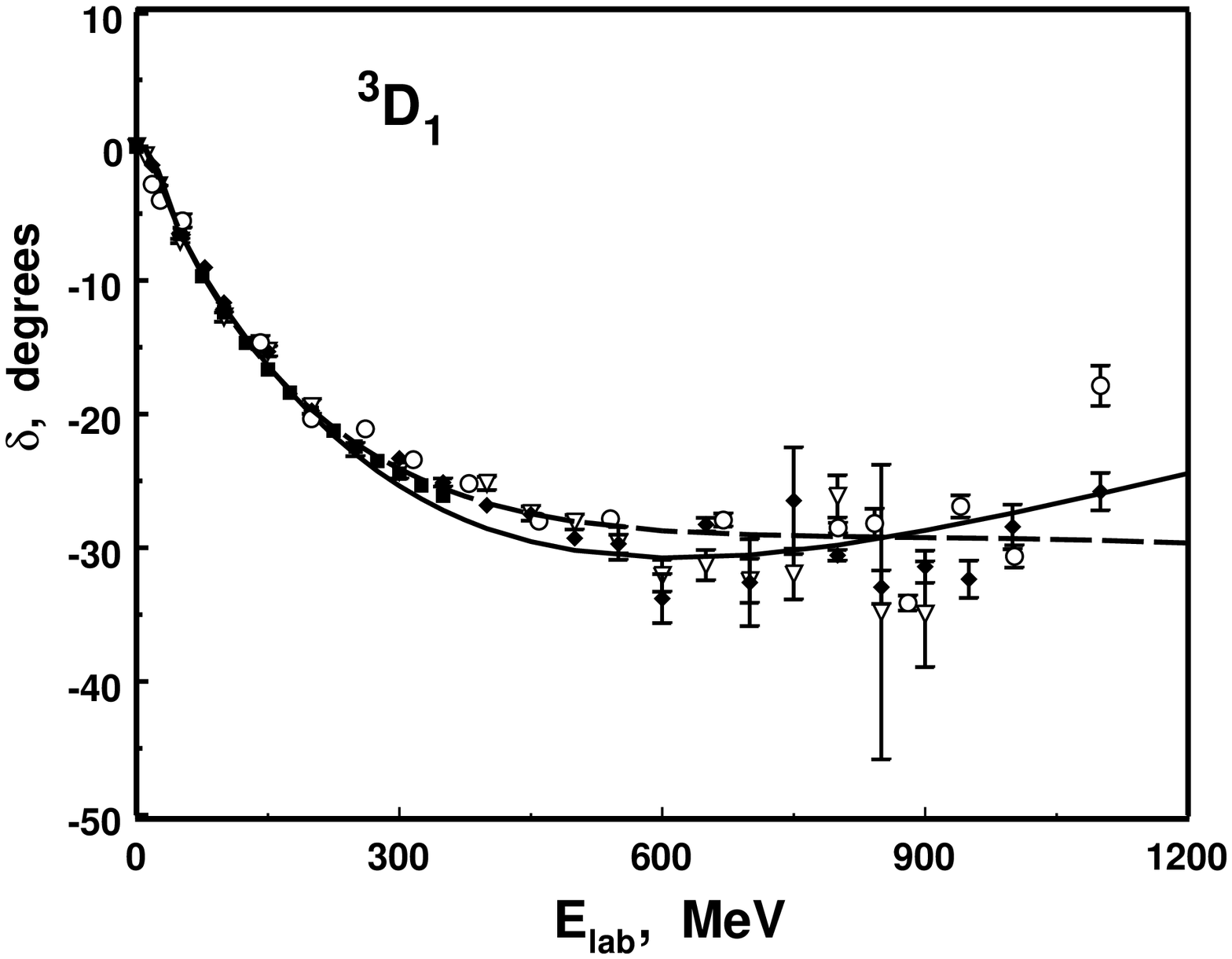}
\epsfysize=50mm \epsfbox{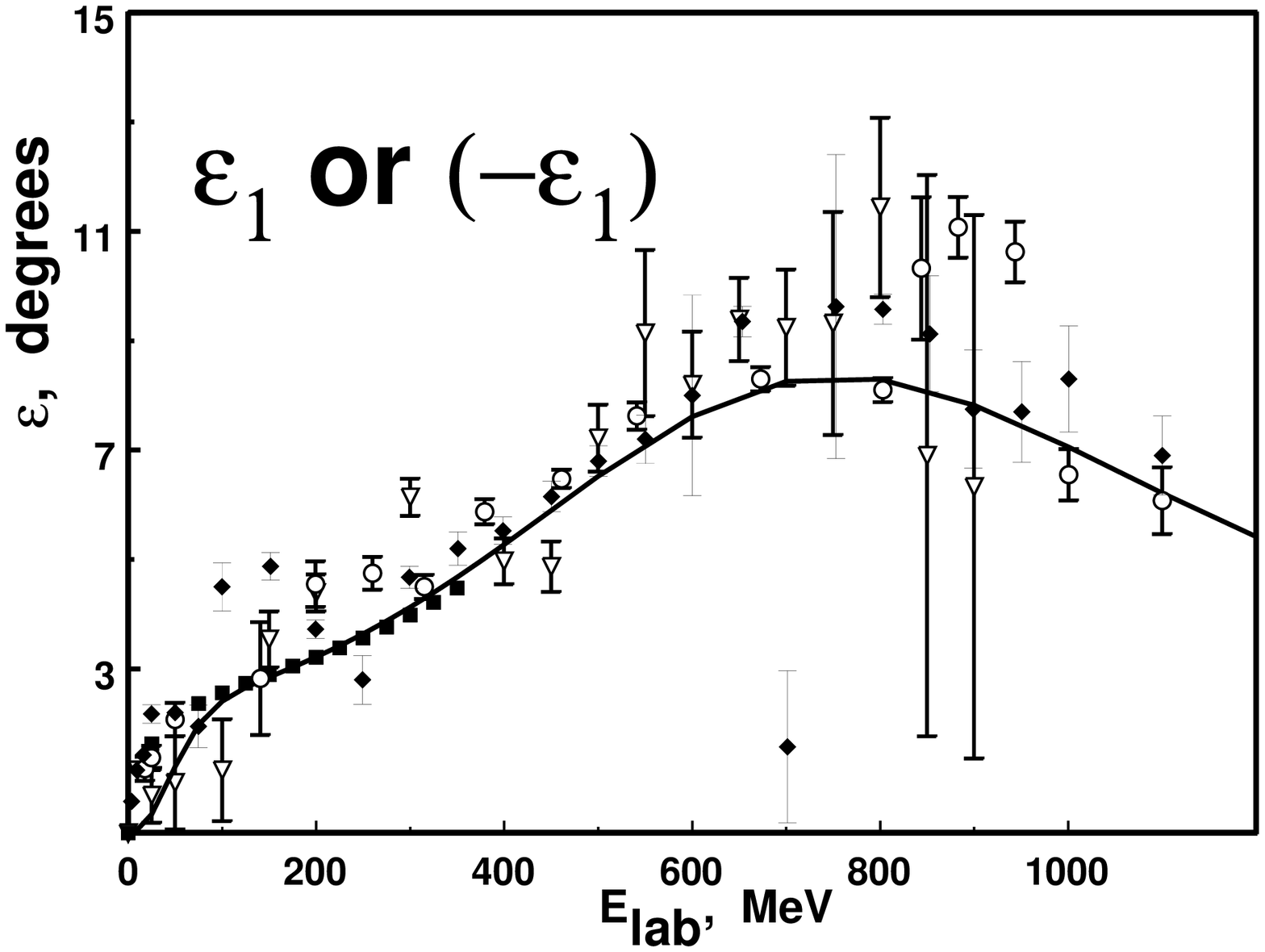}}\caption{Phase shifts and mixing
parameter. Solid line for reconstructed Moscow potential, dashed
lines for reconstructed repulsive core potential. The phase shift
analysis data are from \cite{DataScat}. For $S$ waves there are
two sets of the phase shifts. First one is the original data set
from \cite{DataScat} - small symbols. These data are described by
the repulsive core potential. Second one are the same phase shifts
raised 180 degrees up - big symbols. These data are described by
the Moscow potential. To leave the $S$-matrix unchanged we must
then change the sign of the mixing parameter $\epsilon_1$ for the
Moscow potential. The mixing parameters for both kinds of
potentials differ only by sign in our calculation and
corresponding curves coincide in this figure.}
\end{figure}

\end{document}